\documentstyle[preprint,aps]{revtex}
\begin{document} 
\draft

%To include PS files
\input epsf
\epsfverbosetrue

\newcommand{\ket}[1]{|#1\rangle}
\newcommand{\bra}[1]{\langle #1|}
\newcommand{\braket}[2]{\langle #1|#2\rangle}

\title{A Study of Degenerate Four-quark states in SU(2) Lattice
Monte Carlo}
\author{ A.M. Green\thanks{E-mail address: green@phcu.helsinki.fi} ,
J.~Lukkarinen\thanks{E-mail address: jani.lukkarinen@helsinki.fi},
P. Pennanen\thanks{E-mail address: ppennane@phcu.helsinki.fi}}
\address{ Research Institute for
Theoretical Physics, University of Helsinki, Finland} 
\author{C. Michael\thanks{E-mail address: cmi@liverpool.ac.uk}} 
\address{DAMTP, University of Liverpool, UK}

\maketitle
 
\begin{abstract}
The energies of four-quark states are calculated for geometries in
which the quarks are situated on the corners of a series of
tetrahedra and also for geometries that correspond to gradually
distorting these tetrahedra into a plane. The interest in tetrahedra
arises because they are composed of {\bf three } degenerate
partitions of the four quarks into two two-quark colour singlets.
This is an extension of earlier work showing that geometries with
{\bf two} degenerate partitions (e.g.\ squares) experience a large
binding energy. It is now found that even larger binding energies do
not result, but that for the tetrahedra the ground and first excited
states become degenerate in energy. The calculation is carried out
using SU(2) for static quarks in the quenched approximation with
$\beta=2.4$ on a $16^3\times 32$ lattice. The results are analysed
using the correlation matrix between different euclidean times and
the implications of these results are discussed for a model based on
two-quark potentials.
\end{abstract}

%\newpage
 
%\setcounter{page}{1}
%\pagestyle{plain}
\section{ Introduction}
\label{intro}
In many-particle systems it is often convenient, or even necessary,
to replace the fundamental two-particle interaction by an effective
interaction, which can be very different from the original. For
example, in metals the repulsive coulomb interaction between the
valence electrons -- because of the presence of the underlying ionic
lattice -- is replaced by an effective interaction that is
attractive. Also in nuclei, the free two-nucleon potential gets
strongly modified by the presence of the other nucleons. In both of
these examples, after this removal of the explicit photon and meson
degrees of freedom, the resultant effective interaction is still
mainly in the form of a {\bf two}-body interaction -- with only a
minor three-body term arising in the nuclear case. However, for a
system of quarks interacting via gluon-exchange -- even though this
basic interaction is that of QCD -- little is known in multiquark
systems about the corresponding effective interquark interaction
after the explicit gluon degrees of freedom have been removed. This
is an important question, if realistic calculations are to be made
for interacting quark clusters e.g.\ as in meson-meson,
meson-nucleon or, eventually, nucleus-nucleus scattering. At
present, the only way to carry out these fundamental calculations is
by using Monte Carlo lattice techniques. Unfortunately, mainly due
to the limitations of present day computers, these calculations are
restricted to clusters of, at most, two or three quarks -- see for
example Ref.\ \cite{Fukugita}. Therefore, a bridge is needed between
lattice calculations involving few quarks and conventional many-body
techniques that can accomodate larger numbers of quarks.

In an attempt to throw some light on the relationship between QCD
and the effective interquark interaction, a series of model
calculations have recently been made \cite{MGP}--\cite{GLPM}. In
these, the energies of four quarks in various configurations (e.g.\
on the corners of a rectangle or tetrahedron) have been calculated
in quenched static SU(2) on a $16^3\times 32$ lattice. The reason
for studying four quarks is because of the possibility for
partitioning these into different colour-singlet groups each
containing two quarks -- a situation not possible with 2 and 3 quark
systems. This can then be considered as a step towards the {\bf
scattering} of quark clusters -- in this case meson-meson
scattering. Hopefully, many of the features of an effective
interquark interaction in SU(2) will be preserved in the more
realistic case of SU(3) -- in the same way as, for example, the
ratios of glue-ball masses/string energy are numerically very
similar in SU(2) and SU(3) \cite{CM1}--\cite{Hunt}.

In Refs.\ \cite{GMP}--\cite{GLPM}, the main feature in the results
is that the strongest interaction between two separate two-quark
clusters occurs when two of the three possible partitions of the
clusters are {\bf degenerate}, or almost degenerate, in energy. This
was first observed in Ref.\ \cite{GMP}, where the four-quark binding
energy $E$ dropped by an order of magnitude, when the four-quark
geometry changed from that of a square [$r\times r$] to a
neighbouring rectangle [$r\times (r\pm a)$] -- where $r$ is the
length of a side of the square in lattice units $a$. For example,
$E(3a\times 3a)=-0.054(1)/a= -90$~MeV compared with $E(3a\times 2a \
{\rm or} \ 4a)\approx -0.006(1)/a$. Subsequent works verified this
observation with other geometries, where the two partitions were not
exactly degenerate -- but more so than for the above rectangles. For
example, in Ref.\ \cite{GMP2} this was achieved by tilting the
rectangles out of the planes defining the underlying lattice, and
also with other non-planar geometries. These latter cases showed
that it was indeed the {\bf energy} degeneracy dictating the size of
the interaction and not a geometrical degeneracy. In other words,
with clusters (13)(24) and interquark potentials $V(ij)$, it is the
degeneracy $V(13)+V(24)\approx V(14)+V(23)$ that matters and not
$|\mbox{\bf r}_1-\mbox{\bf r}_3|+|\mbox{\bf r}_1-\mbox{\bf
r}_3|\approx |\mbox{\bf r}_1-\mbox{\bf r}_4|+|\mbox{\bf
r}_2-\mbox{\bf r}_3|$. For large interquark distances $V(ij)\approx
b_s|\mbox{\bf r}_i-\mbox{\bf r}_j|+c$, where $b_s$ is the string
energy and $c$ an additive constant. In that case, the energy- and
geometrical- degeneracy are the same. However, this is only true for
$r_{ij}\ge 0.5$~fm and in the interesting region dominated by
explicit quark interactions it is the energy degeneracy criterion
that must be used. This is contrary to the assumption made with the
so-called flip-flop model of Ref.\ \cite{flip} that takes the
geometrical degeneracy criterion for all interquark distances.

The reason for studying many different 4-quark configurations (i.e.\
rectangles, linear, non-planar, etc.) is to give a representative
selection of geometrical possibilities that arise in practice when,
for example, two mesons scatter. Any quark model for such a
scattering, when reduced to the same conditions as the lattice
calculation [i.e.\ static, quenched, SU(2)] should give for the
above selection of configurations the {\bf same} values for the
4-quark energies $E$. As said earlier, the representative selection
of 4-quark configurations concentrated on those cases where two of
the three possible partitions were degenerate in energy. However, a
potentially interesting case is that of the tetrahedron, where all
three partitions are now degenerate. This is the main subject of
this paper and is discussed in section 2. For completeness, the
series of configurations shown in figure 1 are treated with $1\le d
\le 5$ and $1\le r\le 6$ -- the tetrahedron being $r=d$. The $r=1$
case is then closely related to the earlier work on squares with
$d=1,\ldots,5$ and serves as a check on those calculations, since
they were done with a set of basis states restricted to two in
number and not three as in the present work.

Since the completion of Refs.\ \cite{GMP}--\cite{GMS}, it has been
suggested \cite{MM} that perhaps more care is needed in extracting
energies and their statistical errors from a series of Monte Carlo
results. Therefore, in section 3, the data is analysed taking into
account data correlations between different euclidean times.

\section{The lattice Monte Carlo calculation }
\label{2} 
 \vskip 0.5 cm

\begin{figure}[htp]
\begin{center}
\mbox{\epsfxsize=400pt\epsfbox{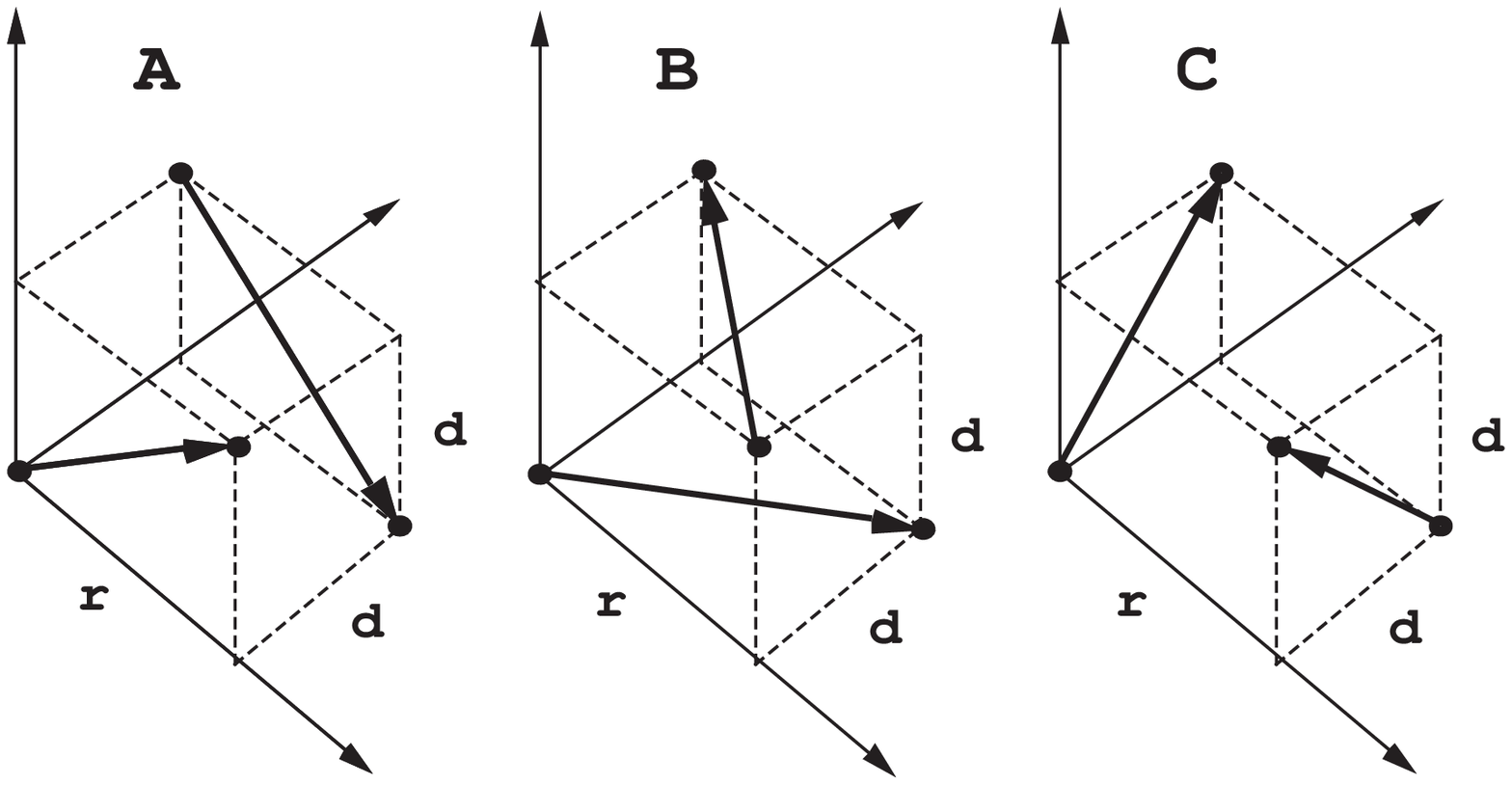}}
\end{center}
\caption{The four-quark geometry based on a square of side $d$
parallel to the $yz$-plane and a distance $r$ from that plane. $A,B$
and $C$ are the three possible partitions. The case when $r=d$ is
called the tetrahedron. \label{f:paths}}
\end{figure}

Since most of the details of the lattice Monte Carlo calculation
have been given in the earlier papers \cite{GMP}--\cite{GMP2}, this
section will only very briefly outline those details concentrating
mainly on aspects specific to the geometry shown in Fig.\
\ref{f:paths}.

Throughout this study the four quarks are treated in the static
quenched approximation on a $16^3\times 32$ lattice in SU(2) with
$\beta=2.4$ -- corresponding to a lattice spacing of $a\approx
0.12$~fm. Both the four-quark energies $V_i(4q)$ and the
corresponding two-quark energies $V_i(2q)$ are extracted on this
lattice. However, the main quantities of interest are the ground
state binding energy $E_{i=1}$ and the excited state energies
$E_{i=2,3}$ defined by the differences
\begin{equation}
\label{Ei}
E_i=V_i(4q)-2V_1(2q)
\end{equation}

As described in Refs.\ \cite{GMP}--\cite{GMP2}, the $V(2q)$
potentials are calculated from a 3*3 variational basis, where --
after fixing the gauge so that all temporal links are set to unity
-- the three states are constructed from basic lattice links that
have been ``fuzzed'' to different levels -- 12, 16 and 20 in the
spatial directions. This fuzzing essentially models a glue flux-tube
between the two quarks in question. For the cases where these two
quarks are not along a given spatial axis [e.g.\ $V(x,y)$], the
appropriate path of links is then constructed as the average of the
two most simple paths connecting $x$ and $y$ -- each consisting of
one straight section along the $x$ and $y$ axes. For $V(4q)$, using
only the maximum fuzzing level of 20, the basis states are
constructed from the different partitions of four quarks into a
product of two two-quark colour-singlet clusters. In some cases it
seems natural to restrict the variational basis to just two
partitions e.g.\ with rectangles the two involving the sides of the
rectangle and not the one constructed from the two diagonal paths.
However, in other cases -- especially for the tetrahedron of
interest here -- it is seems to be necessary to keep all three
partitions.

Having constructed the above paths (e.g.\
$P_i(T_1)=U_i(T_1)\ket{\mbox{Vac}}$ and
$P_j(T_2)=U_j(T_2)\ket{\mbox{Vac}}$) from fuzzy links, Wilson loops
$W_{ij}^T$ -- the quantities from which the final energies are
extracted -- are then simply given, with the present gauge, as the
overlap of paths separated by euclidean time $T=T_2-T_1$ i.e.\
$W_{ij}^T= \braket{P_j(T_1+T)}{P_i(T_1)}$.

The energies $E_i$ in Eq.\ (\ref{Ei}) are expected to be exact in
the sense that they are not the result of a truncated weak or strong
coupling expansion. However, they do contain uncertainties such as:

a) Statistical errors. To a large extent these can be estimated with
reasonable accuracy -- the subject of section 3.

b) Systematic errors. By their very nature they are much harder to
estimate. Throughout, the aim is to attempt at minimizing their
effect. In general, the most obvious source of such errors is the
use of a lattice that is too small (i.e.\ too few points) or has a
grid that is too coarse (i.e.\ $\beta$ too small) for the results to
represent those expected in the continuum limit of the lattice. This
was checked in Ref.\ \cite{GMP2} by using a $24^3\times 32$ lattice
at $\beta=2.5$, which corresponds to a lattice spacing $a\approx
0.082$~fm. In the present type of calculation, since -- as seen from
Eq.\ (\ref{Ei}) -- the signal $E_i$ is a delicate cancellation, it
is necessary to ensure that the appropriate $V_i(4q)$ and $V_1(2q)$
are extracted {\bf simultaneously} from the {\bf same} lattice
configuration. In this way, systematic errors possibly present in
the separate energies $V_i(4q)$ and $V_1(2q)$ are expected to cancel
to some extent.

In order to study different aspects of degenerate geometries, three
separate sets of lattice calculations are performed in this paper.
\begin{itemize} 
\begin{enumerate} 
\item The first concentrates on geometries near to the tetrahedra --
namely -- those cases with $r=d,d\pm 1$ (except for $d=1$, where
only $r=1,2$ are treated). The data is collected into 69 blocks each
containing 32 measurements i.e.\ 2208 measurements in all.
\item The second set of geometries is for $r=1$ and $d=2$ to 5 i.e.\
situations close to that of the squares studied earlier in Refs.\
\cite{GMP}--\cite{GMS}. Here the data is collected into 47 blocks
each containing 64 measurements i.e.\ 3008 measurements in all.
\item In order to see the connection between the present series of
geometries -- especially that in the previous item -- and the
square/rectangle geometry of Refs.\ \cite{GMP}--\cite{GMS}, a third
set of runs is carried out. These are a repeat of the earlier
square/rectangle geometry but with the complete 3*3 basis i.e.\ with
the inclusion of the basis state constructed from the diagonal
paths. Only a limited number of 64 measurements are made -- 4 blocks
each containing 16 measurements. However, this is sufficient to see
how the energies vary in going from the three 2*2 bases (A+B, A+C
and B+C) to the complete A+B+C basis.
\end{enumerate} 
\end{itemize}

\section{Analysis of the data }
\label{3}
As emphasized, for example, in Refs.\ \cite{MM}--\cite{Tous}, care
should be taken when analyzing data generated in Monte Carlo lattice
calculations, in case the data between different euclidean times is
too correlated. This can be taken into account in the following
manner. As discussed in the previous section, the actual quantities
measured are the Wilson loops $W_{ij}^{T}$ between different states
$i,j$ at different time intervals $T$. Here the $i,j$ refer to the
different degrees of fuzzing for the two-quark potentials and, for
the four-quark case, to the different quark partitions. However, the
quantities of interest are the energies ($V$) of these systems. In
Refs.\ \cite{GMP}--\cite{GMS} these were extracted by solving the
eigenvalue equation
\begin{equation}
\label{Eig}
W_{ij}^T a_j^T=\lambda^{(T)}W_{ij}^{T-1} a_j^T,
\end{equation}
where $\lambda_i^{(T)}\rightarrow \exp(-V_i)$ as $T\rightarrow
\infty$. This matrix formulation takes into account mixing between
the different degrees of fuzzing or four-quark partitions, but says
nothing about possible correlations between the different times $T$.
As suggested in Ref.\ \cite{MM} it is convenient to treat the
configuration and time correlations in the same manner. For clarity,
this is illustrated by the specific example of $n$ partitions and $
T$ time differences in the four-quark case. This involves the $(n^2
T)$ matrix elements $W_{11}^T,W_{12}^T,\ldots,W_{nn}^T$, where the
expected equality between, for example, $W_{12}^T$ and $W_{21}^T$ is
not enforced by symmetrizing the data. Let the index
$p=1,\ldots,n^2$ denote these $n^2$ matrices. As discussed in the
previous section, the data ($W_{ij}^T$) in the form of these $(n^2
T)$ matrix elements is collected into $N$ separate blocks, each of
which is already the average of $m$ measurements i.e.\ in all there
are $mN$ measurements of each Wilson loop. The average of these
measurements is denoted by
\begin{equation} 
\label{Av} 
\bar{W}_{ij}^T=\frac{1}{N}\sum^N_{l=1} W_{ij}^T(l).
\end{equation} 
The problem is now reduced to fitting these averages by a
function of the general form 
\begin{equation} 
\label{fun} 
F^T_p=\sum_{k=1}^{k_M}
a_p(k) \exp(-V_kT) 
\end{equation} 
in order to extract the energies $V_k$. As
shown in Ref.\ \cite{MM} this can be achieved by minimizing the
expression 
\begin{equation} 
\label{chi2} 
\chi^2=\sum_{T,T',p,p'}\left(F^T_p-
\bar{W}^T_p\right)M(T,T',p,p') \left(F^{T'}_{p'}-
\bar{W}^{T'}_{p'}\right), 
\end{equation} 
where 
\begin{equation} 
\label{M}
M(T,T',p,p')=NC^{-1}(T,T',p,p') 
\end{equation} 
and $C$ is the covariance matrix
\begin{equation} 
\label{corrn} 
C(T,T',p,p')=\frac{1}{N-1}\sum_{l=1}^N
\left(W_{p}^T(l)- \bar{W}_{p}^T\right) \left(W_{p'}^{T'}(l)-
\bar{W}_{p'}^{T'}\right). 
\end{equation} 
In what follows, it is convenient to
introduce the correlation matrix 
\begin{equation} 
\label{CorrM}
\tilde{C}(T,T',p,p')=\frac{C(T,T',p,p')}
{\sqrt{C(T,T,p,p)C(T',T',p',p')}}
\end{equation} 
The above procedure is stable provided $N$ is large enough. However,
in the present type of lattice calculation this is not guaranteed
and can then lead to $\tilde{C}(T,T',p,p')$ having very small
eigenvalues that produce large fluctuations in $M(T,T',p,p')$. As
argued in Ref.\ \cite{MM}, such eigenvalues correspond to
eigenvectors that alternate in sign as a function of $T$ and so are
not very relevant to smooth fit functions of the type proposed in
Eq.\ (\ref{fun}). It is, therefore, reasonable to remove these
disturbing eigenvalues and here the suggestion made in Ref.\
\cite{MM} is adopted. This simply replaces the smallest eigenvalues
of $\tilde{C}(T,T',p,p')$ by their average. The question then arises
concerning how many $(n_t)$ of the original eigenvalues $(n_T=n^2
T)$ should be retained. Initially a number $n_M$ of these $n_T$
eigenvalues are fixed -- with $n_M$ being chosen as $\approx
\sqrt{N}$ -- a compromise choice found after some experimentation in
Ref.\ \cite{MM}. The average $(\lambda_{av})$ of the remaining
$n_T-n_M$ eigenvalues is then made and those eigenvalues less than
$(\lambda_{av})$ are replaced by $(\lambda_{av})$ itself. Since only
a fraction of the $n_T-n_M$ eigenvalues are replaced in this way,
there still remains $n_t>n_M$ of the original eigenvalues. In
practice the final value of $n_t$ is not strongly dependent on the
original choice of $n_M$. In the present problem $N=69$, $n=2$ or 3
and $ T$ can range from 1 to 5, so that for the largest $\tilde{C}$
matrix only about 8 of the total number of 45 original eigenvalues
should be retained. However, frequently the $T=1$ data are dropped
from the analysis, since they have very small error bars and
presumably contain the effect of several higher energy components
that become negligible already at $T=2$ and so cannot be seen for
$T\geq 2$. Therefore, keeping the $T=1$ data often makes it
difficult to obtain an accurate fit with only a few (1~--~3) terms
$V_i$.

Having arrived at a suitable model for $\tilde{C}$, it remains to
fix the precise form of the function $F^T_p$ in Eq.\ (\ref{fun}),
which is to fit the $n^2T$ matrix elements $\bar{W}^T_p$ by means of
Eq.\ (\ref{chi2}). It is seen that the $F^T_p$ are described by the
$k_M$ potentials $V_k$ and the corresponding $n^2k_M$ amplitudes
$a_p(k)$ i.e.\ in all $(n^2+1)k_M$ parameters. For the four-quark
case the most reliable fits are found with three partitions $(n=3)$
and four time steps $(T=2,\ldots,5)$, which means 36 pieces of data
are to be fitted with $10k_M$ parameters. Therefore, in principle
only three potentials $(V_{1,2,3})$ can at most be extracted.
However, fitting 36 numbers (some having significant error bars)
with 30 parameters ($k_M=3$) would not succeed. In fact, even the
extraction of two potentials using the 20 parameters ($k_M=2$) would
probably not give values of $V_{1,2}$ with a meaningful accuracy. It
is, therefore, necessary to impose some theoretical constraints to
restrict the number ($n^2k_M$) of parameters $a_p(k)$. Since the
data $\bar{W}^T_{i,j}$ involve the overlap of two lattice
configurations $P_i(0)=U_i(0)\ket{\mbox{Vac}}$ and
$P_j(T)=U_j(T)\ket{\mbox{Vac}}$ separated by time $T$, it can be
written as the vacuum expectation
\begin{equation}
\label{Wex}
\bar{W}^T_{i,j}=\bra{\mbox{Vac}} U_j(T)U_i(0)\ket{\mbox{Vac}} 
=\sum_k \bra{\mbox{Vac}}U_j(T)\ket{k}
\bra{k}U_i(0)\ket{\mbox{Vac}}\exp(-V_kT)
\end{equation}

It is, therefore, reasonable to assume that the amplitudes are
\underline{separable} i.e.\ $a_p(k)=a_{ij}(k)=a_i(k)a_j(k)$. This
has two good features:

1) The number of parameters is now reduced to $(n+1)k_M$. Therefore,
fitting the 36 pieces of data from three partitions $(n=3)$ and four
time steps now only requires $4k_M$ parameters. This makes the quite
reliable extraction of two or possibly three potentials feasible.

2) This parametrisation imposes the symmetry $F^T_{ij}=F^T_{ji}$ as
is expected physically. As said earlier this symmetry was not forced
on the original data by some procedure such as taking the average
$0.5(\bar{W}^T_{ij}+\bar{W}^T_{ji})$, since the differences between
$\bar{W}^T_{ij}$ and $\bar{W}^T_{ji}$ can contribute to the error
analysis of the extracted energies.

It should be added that, in the event of two energy states being
degenerate (e.g.\ $V_{k_1}=V_{k_2}$), then
$a_i(k_1)a_j(k_1)+a_i(k_2)a_j(k_2)$ should be replaced by a single
amplitude $a_{ij}$. This is the situation encountered in the
tetrahedron geometry. However, there additional symmetries suggest
the following better model.

In some symmetrical cases, such as four quarks on the corners of a
square or tetrahedron, the number of parameters can be further
reduced by choosing forms of $F^T_p$ that guarantee various
symmetries. For the square, when only the two partitions involving
the sides are considered, the matrix of Wilson loops has the form
\begin{equation}
\label{NVS}
{\bf W}^T=\left(\begin{array}{ll}
W_{11}^T&W_{12}^T\\
W_{21}^T&W_{22}^T\end{array}\right),
\end{equation}                                                         
where not only is the general symmetry $W_{12}=W_{21}$ expected but
also for a square $W_{11}=W_{22}$. In this case Eq.\ (\ref{Eig}) is
easily solved to give for the lowest energy
\begin{equation}
\label{221}
\lambda_1= \frac{W^T_{11}+W^T_{12}}{W^{T-1}_{11}+W^{T-1}_{12}}
\end{equation}              
and for the energy of the first excited state
\begin{equation}
\label{222}
\lambda_2= \frac{W^T_{11}-W^T_{12}}{W^{T-1}_{11}-W^{T-1}_{12}}.
\end{equation}

Therefore, in fitting the data it is reasonable to expect that only
two potentials can be extracted i.e.\ $k_M=2$, and so the fitting
functions must have the form
\[F^T_{1}=F^T_{4}= a(1)\exp(-V_1T)+a(2)\exp(-V_2T) \]
\begin{equation}
\label{FS1}
F^T_{2}=F^T_{3}= a(1)\exp(-V_1T)-a(2)\exp(-V_2T).
\end{equation}
This reduces the number of parameters to four [$V_{1,2}, a(1,2)$] in
order to fit the 16 pieces of data covering four $T$ steps. Here it
should be remembered that the suffix $p$ on $F^T_p$ is fixed as
$p=j+n(i-1)$.

Similarly, for the tetrahedron the matrix of Wilson loops has the
form
\begin{equation}
\label{NVT}
{\bf W}^T=\left(\begin{array}{lll}
W_{11}^T&W_{12}^T&W_{13}^T\\
W_{21}^T&W_{22}^T&W_{23}^T\\
W_{31}^T&W_{32}^T&W_{33}^T
\end{array}\right),
\end{equation}                
where the general symmetries $W_{11}^T=W_{22}^T=W_{33}^T$,
$W_{21}^T=W_{12}^T$, $W_{31}^T=W_{13}^T$, $W_{32}^T=W_{23}^T$ are
expected and, in addition, there are the equalities
$W_{13}^T=W_{12}^T$ and $W_{23}^T=-W_{13}^T$. The "minus" sign
appearing in the last equation is a reminder that the quarks are in
fact fermions even though quarks and antiquarks transform in the
same way under SU(2). This point is discussed in more detail in the
appendix of Ref.\ \cite{GMS}. Therefore, in all, there are only two
independent Wilson loops $W_{11}^T$ and $W_{12}^T$. Again Eq.\
(\ref{Eig}) is easily solved to give for the lowest energy
(occurring twice)
\begin{equation}
\label{331}
\lambda_{1,2}= \frac{W^T_{11}+W^T_{12}}{W^{T-1}_{11}+W^{T-1}_{12}}
\end{equation}              
and for the excited state
\begin{equation}
\label{332}
\lambda_3= \frac{W^T_{11}-2W^T_{12}}{W^{T-1}_{11}-2W^{T-1}_{12}}.
\end{equation}              
In fitting the data it is reasonable to expect that only two terms
can be extracted, i.e.\ $k_M=2$, and so the fitting functions must
have the form
\[F^T_{1}=F^T_{5}=F^T_{9}= a(1)\exp(-V_1T)+a(2)\exp(-V_2T) \]
\begin{equation}
\label{FS2}
F^T_{2}=F^T_{3}=F^T_{4}=F^T_{7}=-F^T_{6}=-F^T_{8}=\\
0.5 a(1)\exp(-V_1T)-a(2)\exp(-V_2T).
\end{equation}
This again reduces the number of parameters to four [$V_{1,2},
a(1,2)$] in order to fit the 12 pieces of data $(n=3)$ covering four
$T$ steps. In this case, due to the high symmetry there are no
longer 36 pieces of independent data.

In the tetrahedral case it is also seen that, when one of the
partitions is removed (i.e.\ $n$ drops from 3 to 2), the Wilson loop
matrix reduces to the same form as that in Eq.\ (\ref{NVS}). As seen
by comparing Eqs.\ (\ref{221}) and (\ref{331}) the lowest eigenvalue
is now the {\bf same} in the $n=2$ and 3 cases. However, the first
excited state is quite different in the two cases -- unlike the
result for most other geometries. For example, as will be seen later
in Table \ref{t:square} of Section \ref{4}, for squares and for
those rectangles so far discussed, the $n=2$ and 3 cases give very
similar results for $E_2$ as well as $E_1$. However, there the third
partition has a higher energy than the other two partitions, whereas
for the tetrahedron all three partitions have the same energy. When
the tetrahedron is distorted into a neighbouring lattice
configuration, it will be seen in Table \ref{tetresults1} of Section
\ref{4} that both $E_1$ and $E_2$ can again be given quite
accurately by $n$=2. It should be added that for very elongated
rectangles the two partitions with the highest energy become more
and more degenerate. In that case $E_1$ is essentially zero and
$E_2$ will depend more and more on the presence of the third
partition.

In an idealised situation, all of the available $n^2$ pieces of data
from $T$=1 to 5 should -- using the above procedure -- be fitted
with a $F^T_p$ containing $k_M\approx n$ potentials. This would then
be the natural extension of Eq.\ (\ref{Eig}) for incorporating into
its $n$ eigenvalues the effects of correlations between different
values of $T$. As it now stands that equation only includes the
mixing between the different fuzzing levels or partitions. However,
in practice, fitting all the $T$ values is not possible and
decisions have to be made concerning both the data
($\bar{W}^T_{ij}$) to actually be fitted and, in the fitting
procedure, the form of $F^T_p$ and also the model for the
correlation matrix in Eq.\ (\ref{CorrM}).

a) In the present problem, the data goes from $T$=1 to 5 and the
number ($n$) of fuzzings (2-quark) or partitions (4-quark) can range
from 1 to 3. For both the 2- and 4- quark systems usually $T$=2 to 5
turns out to be the most suitable range. When the $T$ range is
reduced to 3--5, better values of $\chi^2$ emerge from Eq.\
(\ref{chi2}) but at the expense of larger errors on the extracted
potentials $V_i$. On the other hand, when $T=1$ data is included --
because of the small errors on this new data -- it usually results
in either a ``no-solution'' situation or a $\chi^2$ that is too
large to be meaningful. In addition to the restriction on $T$, a
decision must be made on the number of fuzzings or partitions to be
included. In the 2-quark system only two fuzzings were necessary at
a given time. This is simply a reflection that the different degrees
of fuzzing are effectively very similar and so little is gained by
increasing the number of different fuzzings from 2 to 3. Here the
two largest fuzzing levels (16 and 20) are used i.e.\ $n$ is always
taken to be two in the 2-quark case. In the 4-quark case it is
possible to use all of the data from the different partitions i.e.\
$n$ is always three in the tetrahedron case.

b) In $F^T_p$, once the separable form or the more symmetrical one
in Eqs.\ (\ref{FS1}) or (\ref{FS2}) have been chosen, the only
decision to be made is the value of $k_M$ -- the number of terms
$(V_i)$ to be determined. This is varied between 1 and 3. In the
correlation matrix $\tilde{C}(T,T',p,p')$ of Eq.\ (\ref{CorrM}) the
number of eigenvalues to be initially retained ($n_M$) can also be
varied. Ref.\ \cite{MM} suggests using $n_M\approx \sqrt{N}$, and
this was found to be a suitable value in the four-quark case as can
be seen from Table \ref{t:strategy} (there $N=69$, so that the
$\sqrt{N}$-result is shown as case 11)\@. On the other hand, in the
two-quark system it is necessary to use $n_M = 2$ or 3 to get a
stable fit. Table \ref{t:strategy} shows the effect of varying $n_M$
in a four-quark system (cases 9--13), as well as a fit with no
correlations (i.e.\ $\tilde{C}= {\bf 1}$, case 8)\@. A few general
features can now be seen:

\begin{enumerate}
\item Comparing cases 8 and 11, the use of the correlation matrix
has approximately a one-sigma-level effect on the averages and it
slightly decreases the errors.
\item The exact value of $n_M$ is not important as long as
approximately all the stable eigenvalues are retained (i.e.\
\underline{$n_t$} is in the right range). To estimate this range for
$n_t$ a jackknife procedure on the eigenvalues was carried
out\cite{jk}.
\end{enumerate}

The above choices in the data and fitting procedure are strongly
inter-related and so a strategy is necessary to single out the
optimal set of values for $n$, the range of $T$, $k_M$ and $n_M$. An
example of this is given in Table \ref{t:strategy}. Basically this
strategy amounts to first fitting only a portion of the data e.g.\
$T=$(3 or 4) -- 5, with only a few potentials e.g.\ $k_M=$1 or 2 --
see cases 1--2 in Table \ref{t:strategy}. It is possible to see a
suitable range for $n_t$, and therefore for $n_M$ as well, already
from these fits. Then $k_M$ is increased to find a maximum number of
potentials that can be extracted by using this range of $T$ (see
cases 2,6,7). The data base is then progressively enlarged to
include more timesteps, while trying to keep the value of $\chi^2$
per degree of freedom fixed or decreasing by adding more potentials
if possible (cases 2--7). Naturally, adding a timestep constrains
the potentials more than before, so that the errors of the
potentials decrease accordingly. Initially this works well, but
eventually the $\chi^2$ become larger and larger, so that usually
the results become meaningless before the ultimate stage of
including all of the data (i.e.\ $n=3$ and $T=$1 -- 5). However, it
should be added that the $T=1$ data fitted with $k_M=3$ has to be
treated with caution, if the resulting potentials are significantly
different from those extracted with $T=2 - 5$ and $k_M$=2, since
this may indicate that even higher energy states are polluting the
$T=1$ data.

\vskip 0.5 cm
\section{Results}
\label{4}

The results for the tetrahedron-like geometry in Fig.\ \ref{f:paths}
are shown in Fig.\ \ref{f:energs} and Tables \ref{tetresults1} and
\ref{tetresults2}. Several points arise from Table
\ref{tetresults1}:
\begin{itemize} 
\begin{enumerate} 
\item From earlier work in Refs.\ \cite{GMP}--\cite{GMP2}, for the
corresponding squares [i.e.\ $(d,0)$ in this notation], where two of
the basis states are degenerate, the binding energy of the lowest
state ranges from --0.07 to --0.05 as $d$ goes from 1 to 5 [see
Table \ref{t:square}]. However, now -- even though at least two of
the basis states are always degenerate ($A$ and $B$) -- the ground
state binding energy is always less than that of the corresponding
square [$E_1(d,0)$]. For a fixed $d$, $|E_1(d,r)|$ decreases as $r$
increases. Nothing interesting happens to $E_1$ at $r=d$, at which
point all the basis states are degenerate in energy.
\item For fixed $d$, as $r$ increases from 0 to $d$, the energy of
the first excited state $E_2$ decreases until $E_2(d,d)=E_1(d,d)$.
For $r > d$, $E_2(d,r)$ increases again. This degeneracy of
$E_{1,2}$ for the tetrahedron is a new feature compared with earlier
geometries. As will emerge in the next section, this is a severe
constraint on any model wishing to describe this data.
\item As seen from columns $A+B+C \ (I \ {\rm and} \ II)$, in all
cases the energies extracted using either ($I$) Eqs.\
(\ref{chi2})--(\ref{CorrM}) or ($II$) Eq.\ (\ref{Eig}) are within
the error bars of each other. If a choice has to be made, then the
results $(I)$ are prefered, since they involve a more detailed
analysis of the data.
\item In columns $A+B$ and $B+C$ only the corresponding two basis
states are used. The symbol $S$ means that the extracted $E_1$ are
{\bf exactly} the same as the $A+B+C (II)$ results. In those cases
it is, therefore, sufficient to only use a 2*2 basis. However, the
choice of which 2*2 basis depends on the particular geometry, since
one of these two basis states must have the lowest unperturbed
energy. Since $A$ and $B$ are degenerate in energy, for a given $d$
this amounts to using $A+B$ for $r\leq d$. Whereas, for $r > d$ it
is necessary to use $B+C$, since $C$ now has the lowest unperturbed
energy.
\item Except for the tetrahedra, the values of $E_2$ are essentially
the same in the 2*2 and 3*3 bases.
\item The values of $E_3(d,r)$ are always much higher than
$E_2(d,r)$. However, as discussed in the next section, this second
excited state is dominated by excitations of the gluon field and so
is outside the scope of the models introduced in that section.
\item The last column shows the two 2-quark ground state potentials
$V_1(d,d)$ and $V_1(d,r)$ that enter with this geometry. In the next
section, these are needed for evaluating a model for the 4-quark
binding energies.
\end{enumerate} 
\end{itemize}

Several of the above points are also illustrated in Fig.
\ref{f:energs} for the case of $d/a=3$ and $r/a=0,1,2,3,4$. In
addition, there are shown the theoretical predictions for the $f=1$
limit of the model to be discussed in the next section.

Table \ref{tetresults2} is very similar to Table \ref{tetresults1}
except that $r/a$ is now restricted to unity i.e.\ geometries
nearest to the $(d\times d)$ squares. The following points can be
made:
\begin{itemize}
\begin{enumerate}
\item Again the energies extracted by Eqs.\
(\ref{chi2})--(\ref{CorrM}) ($I$) and Eq.\ (\ref{Eig}) ($II$) are
within the error bars of each other -- with ($I$) being preferable,
since it involves a more detailed analysis of the data.
\item For comparison in $E_i(d\times d)$ are given the energies of
the nearby squares from Ref.\ \cite{GMP2} with only a 2*2 basis but
analysed with method $(I)$. In general, these all have a somewhat
larger binding energy, indicating that the binding energy is
maximized for the square geometry.
\end{enumerate}
\end{itemize}

In Table \ref{t:square} results are given for various
square/rectangle geometries. The main purpose of this set of runs is
to see any change in the results for $E_{1,2}$ extracted from the
2*2 bases, when this is extended to a 3*3 basis by including a third
state $C$. Also an estimate can now be made for the position of the
second excited state $E_3$.

Several comments should be made about these results
\begin{itemize}
\begin{enumerate}
\item For the lowest energy $E_1$, introducing the third basis state
has little effect on the results given by the two basis state
calculation provided the partition with the lowest energy is one of
those two states in that basis. In fact, for squares (i.e.\ $r=d$)
the results are identical for the $A+B+C$ and $A+B$ bases. However,
in a general case when using a 2*2 basis, for $d >r$ it is necessary
to include state $A$ and for $d<r$ state $B$. Remember that in this
notation it is state $C$ that is constructed from the diagonal
two-quark colour singlets and so always has the highest unperturbed
energy.
\item The energy $E_2$ of the first excited state is less dependent
on the choice of basis -- with the 3*3 and 2*2 possibilities giving
in most cases essentially the same results.
\item From column $A+B+C$ it is seen that $E_3$ -- the energy of the
second excited state -- is always considerably higher in energy than
$E_2$.
\item The $E_i(P)$ column is the result of many more measurements
than the following columns. Therefore, in view of the facts in the
previous items, these numbers should be considered as the most
accurate for rectangles. They are basically the same as those in
refs.\ \cite{GMP1}, \cite{GMP2} and \cite{GMS}, except that the data
has been subjected to the improved analysis ($I$) using Eqs.\
(\ref{chi2})--(\ref{CorrM}).
\end{enumerate}
\end{itemize}

\section{An interpretation of the results} 
\label{5} 
One of the main reasons for embarking on this work is the attempt to
find a model which gives a simple understanding of the four-quark
binding energies in terms of the corresponding two-quark potentials.
At first sight, it may seem that the best such model should be able
to explain all of the energies extracted from the lattice
calculation of the previous section. However, this would be not only
too ambitious but also it would be outside of the goal of the
desired model. When the lattice data is analysed by directly
diagonalizing Eq.\ (\ref{Eig}), then the number of energies
extracted is equal to the number of basis states i.e.\ three for the
tetrahedral geometry. On the other hand, when the expression in Eq.\
(\ref{chi2}) is minimized, the number of extracted energies depends
very much on the quality of the lattice data. For the tetrahedron,
as seen in Tables \ref{tetresults1}--\ref{tetresults2}, it is also
possible to extract the three lowest eigenvalues and, in principle,
even more could be obtained. This point is discussed in more detail
in ref.\ \cite{ext}, where also the results from other four-quark
geometries are extracted using Eqs.\ (\ref{chi2})--(\ref{Wex}).
However, for the model to be discussed below only the lowest {\bf
two} lattice eigenvalues are of interest, since the third is
dominated by excited gluon components. Such a feature is outside any
model that incorporates an interaction that is based on the
two-quark potential in its {\bf ground} state. It can be seen that
the third state is basically a gluonic field excitation as follows:

1) As shown in the above tables and also in the appendix of Ref.\
\cite{GMS}, except for the tetrahedra, the addition of a third basis
state into the lattice calculation has only a minor effect on the
results using only two basis states -- indicating that the structure
of the third eigenstate is very different from the lower two states.

2) In the linear case, on a lattice where all three partitions are
constructed from links along a single spatial axis, i.e.\ without
the need for introducing combinations of links along different axes,
these partitions are linearly dependent leading to a singular
matrix, if a three basis calculation is attempted in Eq.\
(\ref{Eig}).

The previous paragraph has shown that only the lowest two lattice
eigenvalues are of interest when constructing a model based on the
basic two-quark potential in its ground state. To this end, in
refs.\cite{GMP}--\cite{GMS} it was proposed that the lattice
energies $E_i$ should be fitted by a model defined in terms of only
two basis states $A$ and $B$. In this model the $E_i$ are given as
the eigenvalues of the equation
\begin{equation}
\label{Hamf}
\left[{\bf V}(f)-\lambda_i(f) {\bf N}(f)\right]\Psi_i=0,
\end{equation}
with
\begin{equation}
\label{NVf}
{\bf N}(f)=\left(\begin{array}{ll}
1&f/2\\
f/2&1\end{array}\right)\ \ {\rm and}\ \ 
{\bf V}(f)=\left(\begin{array}{cc} v_{13}+v_{24} & fV_{AB}\\
fV_{BA}&v_{14}+v_{23}\end{array}\right),          
\end{equation}
where $\lambda_i=E_i+v_{13}+v_{24}$ -- the configuration
$A=(q_1\bar{q}_3)(q_2\bar{q}_4)$ being the lowest in energy of the
three possible partitions into two two-quark singlets. The
off-diagonal matrix element
\begin{equation}
\label{AVB}
\bra{A}V\ket{B}=V_{AB}=V_{BA}=
\frac{1}{2}\left(v_{13} +v_{24} +v_{14}+v_{23} 
- v_{12}-v_{34} \right)
\end{equation}
is of the form expected from a two-quark isovector potential
\begin{equation}
\label{vcol}
V_{ij}=-\frac{1}{3} {\bf \tau}_i\cdot{\bf \tau}_j v_{ij}.          
\end{equation}
The additional factor of $f$ appearing in the off-diagonal matrix
elements is to be interpreted as a gluon field overlap factor. In
the weak coupling limit $f$ is unity. However, in general this limit
is found to result in too much binding compared with the lattice
results. Therefore, $f$ is treated as a phenomenological factor,
which is adjusted to fit the lattice energies. The problem is then
reduced to understanding the resulting values of $f$, which have
essentially simply replaced the lattice energies. That this is a
reasonable model is supported by several points:
\begin{itemize}
\begin{enumerate}
\item This factor $f$ is approximately unity when all four quarks
are close together. This is indeed where the $f=1$ weak coupling
limit should be best. 
\item For a given geometry, a single value of $f$ qualitatively
explains quite well both $E_1$ and $E_2$.
\item When $f$ is parametrized as
\begin{equation}
\label{param}
f=\exp[-k b_s S],
\end{equation}
where $S$ is the minimal area of a surface bounded by the straight
lines connecting the four quarks and $b_s a^2$ is the string energy
density with the value $\approx 0.070$, then it is found that -- for
squares and rectangles -- the parameter $k$ is reasonably constant
at $\approx 0.50\pm 0.05$. Of course, the ultimate goal is to find
such a parametrization in which the corresponding parameter(s) $k$
would be strictly constant for all geometries (i.e.\ squares,
tetrahedra,$\ldots$). If this selection of geometries is
sufficiently representative, then the reasonable assumption would be
that this same parametrization should work for other geometries not
calculated on the lattice.
\end{enumerate}
\end{itemize}             
Prior to this work on tetrahedrons the geometries considered had, at
most, two of the three possible partitions being degenerate in
energy (e.g.\ for squares). In these cases, as seen from Table
\ref{t:square}, it is found that the lattice energies $E_1$ and
$E_2$ are essentially the same for the three-basis-state calculation
$(A+B+C)$ and those two-basis-state calculations ($A+B,\ A+C$ and
effectively $B+C$) which involve the basis state with lowest
unperturbed energy. This is one of the reasons why the 2*2 version
of the $f-$model in Eq.\ (\ref{NVf}) was quite successful for a
qualitative understanding of these cases. However, for tetrahedra
and the neighbouring geometries calculated in Section 2, it now
seems plausible to extend the $f$-model to the corresponding 3*3
version in which
\begin{equation}
\label{NVf3}
{\bf N}(f)=\left(\begin{array}{lll}
1&f/2&f'/2\\
f/2&1&-f''/2 \\
f'/2&-f''/2&1\end{array}\right)\ \ {\rm and}\ \ 
{\bf V}(f)=\left(\begin{array}{ccc}
v_{13}+v_{24} & fV_{AB}& f'V_{AC}\\
fV_{BA}&v_{14}+v_{23}&-f''V_{BC} \\
f'V_{CA}&-f''V_{CB}&v_{12}+v_{34}\end{array}\right),          
\end{equation}
where the negative sign in the BC matrix elements is of the same
origin as the one in Eq.\ (\ref{NVT}). This extension has both good
and bad features. On the positive side, all three basis states are
now treated on an equal footing. This is convenient when considering
some general four-quark geometry, since it is then not necessary to
choose some favoured 2*2 basis, which could well change as the
geometry develops from one form to another. On the negative side, in
the weak coupling limit (i.e.\ $f,f',f''\rightarrow 1$) the 3*3
matrix of Eq.\ (\ref{Hamf}) becomes singular -- in the sense that
adding columns $B$ and $C$ results in column $A$. However, in this
limit, each of 2*2 matrices corresponding to the three possible
partitions A+B, A+C and B+C gives the same results. Away from weak
coupling the 3*3 matrix is no longer singular, but now the three
possible 2*2 partitions do not necessarily give the same results.
Below an attempt is made to minimize the differences between these
three partitions, since in the corresponding lattice calculation the
differences in most cases are indeed small.

The strategy of trying to mimic the lattice calculation by means of
the $f$-model in Eqs.\ (\ref{Hamf}) and (\ref{NVf}) is carried out
at two different levels.
\begin{itemize}
\begin{enumerate}
\item The lattice calculation is made in the static quenched
approximation with the SU(2) gauge group. Therefore, in the
$f$-model there must not be any kinetic energy term (i.e.\ static
quarks) and also quark-antiquark pair creation (for example in the
form of meson exchange between quark clusters) must not be included
(i.e.\ the quenched approximation). Furthermore, the two-quark
potential in Eq.\ (\ref{vcol}) is expressed in terms of
$\tau$-matrices -- the generators of SU(2).
\item The lattice calculation, as said earlier, gives essentially
the same results for any of the three partitions A+B, A+C and B+C as
the complete A+B+C basis. As will be seen below, this point is more
difficult to mimic. Since this feature is automatically encoded in
the matrix elements of Eq.\ (\ref{NVT}), one possibility for the
$f$-model is to attempt to mimic in more detail the form of these
matrix elements.
\end{enumerate}
\end{itemize}             

For the tetrahedron geometry, the $f$-model as written in Eq.\
(\ref{NVf3}) takes on a particularly simple form since each of the
$v_{ij}$ are equal and also $f=f'=f''$. In this case the eigenvalues
are
\begin{equation}
\label{eigvs}
E_1=E_2=E_3=0.
\end{equation} 
This has the positive feature that $E_1$ and $E_2$ are degenerate as
in the lattice results of Table \ref{tetresults1}. However, they are
degenerate at {\bf zero} binding energy. This latter feature is
unavoidable for the model in its present form, since there is only
one energy scale present. This is the two-quark potential between
each of the quarks, which is, of course, independent of the pair of
quarks chosen, since all interquark distances are the same for the
tetrahedron. It is, therefore, necessary to introduce a second
energy scale into the model. However, any improvements in the model
have very limited choices, since there are only two different matrix
elements involved -- the diagonal ones all equal to $-E$ and the
off-diagonals ones all equal to $\pm0.5fE$. Therefore, the most
general modifications are to change the diagonal matrix elements to
$d_1-E$ and the off-diagonal ones to $\pm 0.5f(d_2-E)$. This results
in the eigenvalues
\begin{equation}
\label{eigvs2}
E_1=E_2=\frac{d_1+0.5fd_2}{1+0.5f} \ \ {\rm and} \ \
E_3=\frac{d_1-fd_2}{1-f}.
\end{equation}
At first sight it may appear that there is sufficient information to
now extract the new parameters $d_{1,2}$, since $f$ can be estimated
using the parameters (assumed to be universal) from other geometries
-- thus leaving two equations for $E_{1,2}$ and $E_3$ and the two
unknowns $d_{1,2}$. However, as said before, this is too much to
demand from the $f$-model, since in the lattice calculation the
third basis state in the complete A+B+C basis generally plays a
minor role in determining the values of $E_{1,2}$ and, therefore,
the third eigenvalue is presumably dominated by an excitation of the
gluon field. Even so, it is of interest to see that a similar
feature now arises with $E_3$ in Eq.\ (\ref{eigvs2}), since this
third state is removed in the weak coupling limit i.e.\
$E_3\rightarrow \infty$ as $f\rightarrow 1$. However, the $f$-model
-- if it is to be successful -- should only be expressed in terms of
the lowest energy gluon configurations, since the gluon field is not
explicitly in its formulation, but only appears {\bf implicitly} in
the form of the two-quark potentials and the $f$-factors. In view of
this, no quantitative attempt should be made to identify the second
excited state emerging from the lattice calculation as $E_3$ in the
$f$-model.

In Ref.\ \cite{jens} it was shown that the {\bf two}-state model of
Eqs.\ (\ref{NVf}) with the overlap factor $f=1$ agreed with
perturbation theory upto fourth order in the quark-gluon coupling
[i.e.\ to $O(\alpha^2)$] and gives $E_{1,2}$=0 for tetrahedra.
Therefore, the non-zero lattice results for small tetrahedra must be
of $O(\alpha^3)$ at least. Another aspect of this special situation
for tetrahedra is also seen -- when extracting or interpreting the
value of $E_2$ -- by the need for the third basis state both in the
lattice calculation (see Table \ref{tetresults1}) and in the
$f$-model, since in comparison with Eq.\ (\ref{eigvs2}) the two
basis state version gives
\begin{equation}
\label{eigvs3}
E_1=\frac{d_1+0.5fd_2}{1+0.5f} \ \ {\rm and} \ \
E_2=\frac{d_1-0.5fd_2}{1-0.5f}
\end{equation}
i.e.\ both the two- and three- basis state models have the same
ground state, but the latter does not show the $E_1=E_2$ degeneracy.

In Fig. \ref{f:energs}
the predictions are shown for the $f=1$ limit of the above model in
the case of $d=3$. If, in the notation of Fig. 
\ref{f:paths}, the appropriate two
potentials in state $A$(or $B$) and state $C$ are defined as
$v_a=V_1(3,r)$ and $v_c=V_1(3,3)$, respectively, then 
\begin{equation} 
\label{f11}
V_1(4q)=\frac{8v_a-2v_c}{3} \ \ {\rm and} \ \ V_2(4q)=2v_c. 
\end{equation} 
Here the $V_i(4q)$ are defined in Eq.\ (\ref{Ei}). To extract $E_i$
for $r\leq d$ the two-quark potential $V_1(2q)$ is taken to be
$v_a$, whereas for $r>d$ the appropriate potential is $V_1(2q)=v_c$.

The expressions in Eq.\ (\ref{eigvs2}) are not particularly useful
unless there is a model for the parameters $d_{1,2}$. However, since
it's not the purpose of this paper to make a comprehensive study of
models covering all the 4-quark geometries considered in earlier
works \cite{GMP1}--\cite{GLPM}, only a few general remarks will be
made here for the tetrahedron geometry. Models for the $d_{1,2}$
need extensions of the potential in Eq.\ (\ref{vcol}), so that for
the tetrahedron there are two energy scales present. Here several
ways of achieving this goal are suggested:
\begin{itemize}
\begin{enumerate}
\item The effect of an isoscalar two-quark potential.

As discussed in Ref.\ \cite{GMP1}, an isoscalar potential $w_{ij}$
can be introduced into $V_{ij}$ -- ensuring $V_{ij}=v_{ij}$ for a
colour singlet two-quark system -- by the form
\begin{equation}
\label{vcol1}
V_{ij}=-\frac{1}{3} {\bf \tau}_i\cdot{\bf \tau}_j 
\left(v_{ij}-w_{ij}\right) +w_{ij} .         
\end{equation}
In this case, $d_1=d_2=4w$ since all of the $w_{ij}$ are now equal
to $w$ and results in $E_1=E_2=E_3=4w$. Therefore, from Table
\ref{t:square} it is seen that the $w$ range from --0.0035 to
--0.0070 i.e.\ they have values much smaller than the corresponding
$v_{ij}=v$ given in the last column. A similar feature was found in
Ref. \cite{GMP1}, when the form in Eq.\ (\ref{vcol1}) was introduced
to improve the model fit for squares and rectangles. However, as
shown in Ref.\ \cite{jens}, in perturbation theory all terms of
$O(\alpha^2)$ are included in the two state model of Eq.\
(\ref{NVf}) with $f=1$. Therefore, in the weak coupling limit
$w_{ij}$ must be of $O(\alpha^3)$ at least.

\item The effect of a three- or four-body potential.

The $f$ factor is itself a four-body operator. However, it is
conceivable that additional multiquark effects arise. Some
perturbative possibilities are discussed in Ref.\ \cite{jens}. There
it is shown that all three-quark terms arising from three gluon
vertices always vanish, but that the four gluon vertex can
contribute to 2-,~3- and 4-quark terms at $O(\alpha^3)$. However, in
the tetrahedral case ($r=d$), cancellations result in this
particular 4-quark term also vanishing.
\item The effect of non-interacting three gluon exchange processes.

These are also discussed qualitatively in Ref.\ \cite{jens} and
contribute at $O(\alpha^3)$ to 2-,~3- and 4-quark potentials.
\item The effect of two quark potentials in which the gluon field is
excited.

The first excited state [$V_1^*(r)$] of the two-quark potential
$V_1(r)$ is approximately given by $V_1^*(r)\approx V_1(r)+\pi /r$
-- see for example refs. \cite{CM1,CM2}. Therefore, if a fourth
state -- based on such an excited state -- is introduced into the
model, since it is higher in energy than the three degenerate basis
states so far considered, it will give attraction in the ground
state. Furthermore, as the size of the tetrahedron increases this
fourth state will approach the other three states, so that the
attraction felt in the ground state will increase -- a trend seen in
the tetrahedron results for $E_{1,2}$ in Table \ref{tetresults1}.

\end{enumerate}
\end{itemize}             

The above isoscalar potential option now offers a reason for
$E_1=E_2\not =0$. But, unfortunately, $E_3$ is still equal to
$E_{1,2}$ since $d_1=d_2$. However, there is no reason to expect any
three or four body forces to also be purely isoscalars. In this
case, their contributions to $d_1$ and $d_2$ could be different and
through the presence of the $1-f$ factor in Eq.\ (\ref{eigvs2}) any
estimates of $E_3$ could be very model dependent. \vskip 0.5 cm

\section{Conclusions}
\label{6}
\begin{figure}[p]
\caption{The binding energies -- in units of the lattice spacing --
of the four-quark states for the geometry of Fig.\
\protect\ref{f:paths} for $d=3$ and
$r=0,1,2,3,4$. \protect\\ Solid lines show lattice results:
\protect\\ $\diamond$ -- the ground-state binding energy $E_1$.
\protect\\ $\times$ -- the first excited state energy $E_2$.
\protect\\ Dashed and dotted lines show model results with $f=1$ 
from Eq.\ (\protect\ref{f11}) for $E_1$ and $E_2$ respectively.
\label{f:energs}}
\begin{center}
\mbox{\epsfxsize=420pt\epsfbox{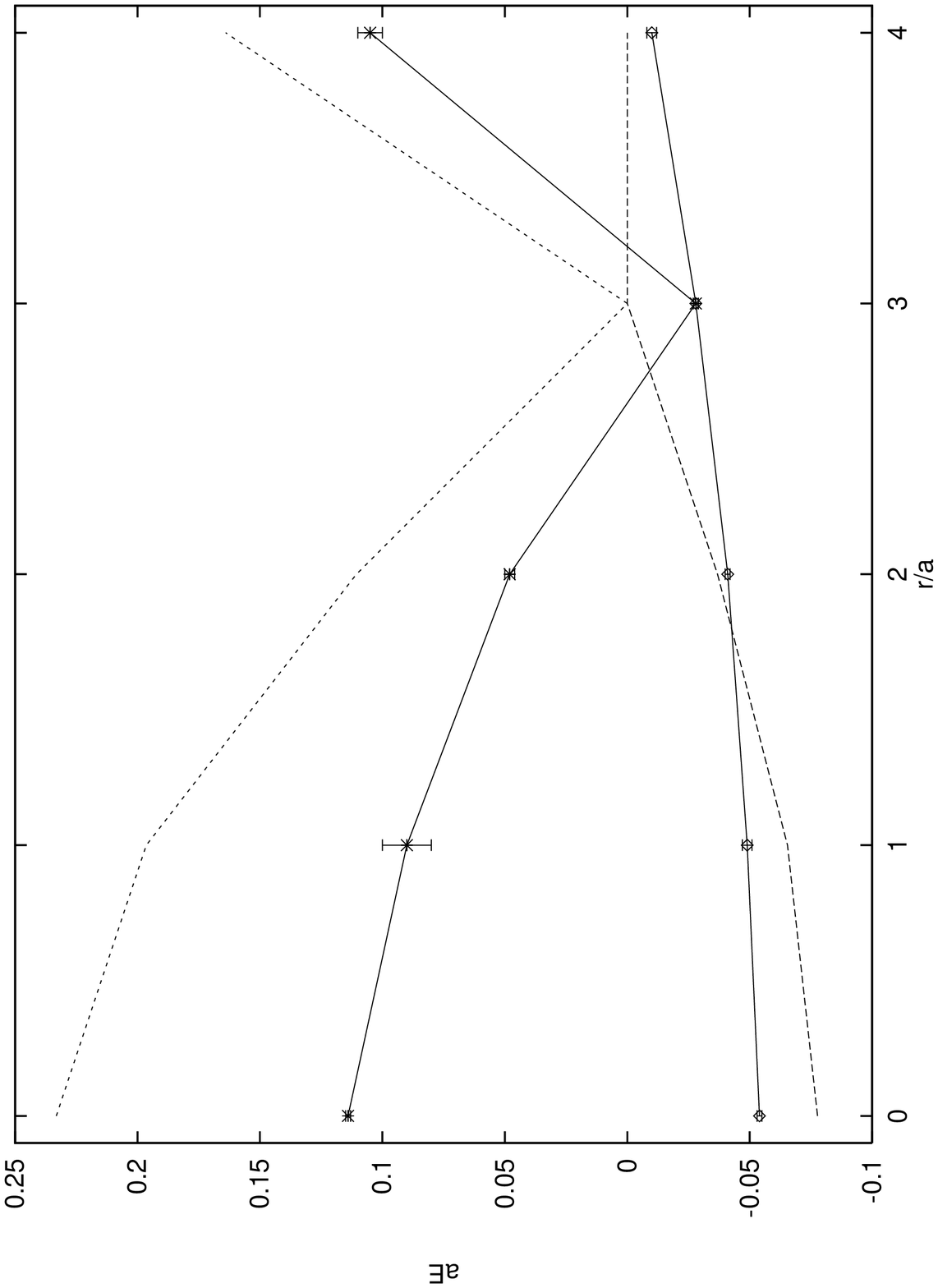}}
\end{center}
\end{figure}

This paper discusses three separate aspects of four-quark energies:  

1) Four-quark configurations (tetrahedra) involving {\bf three}
degenerate basis states.\\ This showed the interesting result that
-- for tetrahedra -- the ground and first excited states are
degenerate in energy. The onset of this degeneracy can be seen in
Fig. \ref{f:energs}, where -- as a square ($r=0$) gets deformed into
a tetrahedron ($r=d=3$) -- the ground and first excited states,
originally at --0.053 and 0.116, become degenerate at --0.026. This
is a new feature not observed in earlier four-quark configurations
(e.g.\ squares, linear,$\ldots$).

2) The use of the correlation matrix in euclidean time for
extracting energies from the basic lattice data.\\ Because the
four-quark binding energy is the difference of two quantities that
are comparable in magnitude, the final result involves a rather
delicate cancellation -- see Eq.\ (\ref{Ei}). It was, therefore,
considered worth while to investigate the effect of correlations in
the basic data. Here it is shown that such an improved analysis of
the lattice data leads to results that are essentially the same as
those in earlier, less complete, analyses.

3) The construction of a model in an attempt to explain these
energies.\\ Here it is seen that the $f$-model, introduced in
earlier papers, needs to be modified before it can be used to
described the tetrahedron results. Hopefully, a guide to these
modifications can be suggested by perturbation theory, which has
already proven useful in justifying the basic model at small
interquark distances.

\begin{acknowledgements}
The authors wish to thank J.Paton and J.Lang for useful
correspondence. They also acknowledge that these calculations were
carried out at the Helsinki CRAY X-MP. In addition this work is part
of the EC Programme ``Human Capital and Mobility'' -- project number
ERB-CHRX-CT92-0051.
\end{acknowledgements}

\begin{table}[p]
\caption{The binding energies $E_k(d=3a,r=4a)$ obtained with method
(I) from Eqs.\ (\protect\ref{chi2})--(\protect\ref{CorrM}).
\protect\\ $k_M$ is the number of energies extracted. \protect\\ $T$
is the time range analysed. \protect\\ $n_M$ is the number of
eigenvalues initially retained. \protect\\ $n_t$ is the actual
number of unchanged eigenvalues finally retained using $n_M$.
\protect\\ $\chi^2$/DoF is the ratio of Chi-squared/Degrees of
Freedom with the actual result shown in the parentheses. \protect\\
Case 8 is a fit with a unit correlation matrix and is equivalent to
the usual Least Squares fit.\protect\\ \label{t:strategy}}

\newpage
%\renewcommand{\baselinestretch}{1.0}
%\Huge
%\large
\begin{center}
\begin{tabular}{c|cc|cc|ccc|c}
Case & $k_M$ & $T$ & $n_M$ & $n_t$ & \multicolumn{3}{c|}{$E_k$} &
$\chi^2$/DoF \\ \hline\hline
1 & 2 & 4--5 & 12 & 15 & --0.06(3) & 0.10(4) & & 230/10 (23) \\
\hline\hline
2 & 3 & 4--5 & 12 & 15 & --0.03(3) & 0.07(4) & 0.4(2) & 3.9/6 (0.6)
\\ \hline
3 & 3 & 3--5 & 10 & 16 & --0.009(6) & 0.097(8) & 0.37(3) & 8.9/15
(0.6) \\ \hline
4 & 3 & 2--5 & 8 & 17 & --0.009(2) & 0.105(2) & 0.400(7) & 12/24
(0.5) \\ \hline
5 & 3 & 1--5 & 8 & 16 & --0.003(2) & 0.116(2) & 0.424(2) & 100/33
(3.0) \\ \hline\hline
6 & 4 & 2--5 & 8 & 17 & --0.010(3) & 0.097(7) & 0.40(3) & 
10/20 (0.5) \\
&&&&& 23.405(2) &&& \\ \hline
7 & 4 & 1--5 & 8 & 16 & --0.004(195) & 0.108(36) & 0.424(95) & 43/29
(1.5) \\
&&&&& 1.8(6.1) &&& \\ \hline\hline
8 & 3 & 2--5 & -- & -- & --0.0100(19) & 0.1035(25) & 0.393(14) &
8.8/24 (0.4) \\ \hline
9 & 3 & 2--5 & 0 & 7 & --0.0091(19) & 0.1045(23) & 0.398(10) & 10/24
(0.4) \\ \hline
10 & 3 & 2--5 & 2 & 11 & --0.0090(17) & 0.1053(24) & 0.399(7) & 
10/24 (0.4) \\ \hline
11 & 3 & 2--5 & 8 & 17 & --0.0090(17) & 0.1054(23) & 0.400(7) & 
12/24 (0.5) \\ \hline
12 & 3 & 2--5 & 15 & 22 & --0.0091(17) & 0.1054(22) & 0.400(7) &
16/24 (0.7) \\ \hline
13 & 3 & 2--5 & 36 & 36 & --0.0087(17) & 0.1065(19) & 0.400(5) &
28/24 (1.2) \\ \hline\hline
14 & \multicolumn{4}{c|}{method (II)} & --0.010(2) & 0.105(5) &
0.39(3) & -- \\
\end{tabular}
\end{center}
\end{table}

\begin{table}[p]
\caption{The binding energies $E_i$ of four quarks in the geometry
of Fig.\ \protect\ref{f:energs} for $d=1$ with $r=1,2$ and
$d=2,\ldots,5$ with $r=d, d\pm 1$. Here $r=d$ are the tetrahedra.
\protect\\
1) $A+B+C$ (I) is from Eqs.\
(\protect\ref{chi2})--(\protect\ref{Wex}) with $A,B,C$ shown in
Fig.\ \protect\ref{f:energs}. \protect\\
2) $A+B+C$ (II) is from Eq.\ (\protect\ref{Eig}) in a 3*3 basis --
see Ref.\ \protect\cite{GMS}. \protect\\
3) $A+B$ and $B+C$ are from Eq.\ (\protect\ref{Eig}) using only 2*2
bases \protect\\
4) $V_1$ are the two-quark potentials i) $V_1(d,d)$ and 
ii) $V_1(d,r)$  \protect\\
5) The symbol S indicates that the entry in the table is the 
same as that to its left -- within numerical accuracy 
i.e.\ rounding errors. \protect\\
These results are from 2208 measurements contained in 69 blocks
of 32 measurements each. \protect\\ \label{tetresults1}}

\newpage

\renewcommand{\baselinestretch}{1.0} 
\Huge 
\large
\begin{center} 
\begin{tabular}{cc|c|c|c|c||r@{\hspace{0.1cm}}l} 
$(d/a,r/a)$&$E_i$&$A$+$B$+$C$ (I)&$A$+$B$+$C$ (II)&
$A$+$B$&$B$+$C$&\multicolumn{2}{c}{$V_1$}\\ \hline 
(1,1)&$E_1$&--0.0145(4)&--0.016(2) &S&S&i)&0.4885(1) \\
&$E_2$&"&"&--0.016(2)&--0.016(2) && \\
&$E_3$&0.85(2)&0.834(4)&&&& \\ \hline 
(1,2) &$E_1$&--0.0034(4)&--0.003(1)&0.03(1)&S&i)&0.4885(1) \\
&$E_2$&0.262(2)&0.265(2)&0.265(2)&0.262(2)&ii)&0.6023(3)\\
&$E_3$&0.95(9)&0.849(3)&&&& \\ \hline 
(2,1)&$E_1$&--0.0453(9)&--0.043(2)&S&--0.042(2)
&i)& 0.6689(4) \\
&$E_2$&0.083(2)&0.085(2)&0.086(2)&0.084(2)
&ii)&0.6023(3) \\
&$E_3$&0.78(7)&0.67(3)&&&& \\ \hline 
(2,2)&$E_1$&--0.0202(8)&--0.020(1)&S&S&i)&0.6689(4) \\
&$E_2$&"&"&--0.017(2)&--0.017(2) &&\\
&$E_3$&0.524(6)&0.49(3)&&&& \\ \hline 
(2,3)&$E_1$&--0.010(2)&--0.008(2)&--0.002(4)&--0.008(1) 
&i)&0.6689(4) \\
&$E_2$&0.147(3)&0.147(2)&0.147(2)&0.150(5)&ii)&0.7421(6) \\
&$E_3$&0.60(3)&0.58(2)&&&& \\ \hline
(3,2)&$E_1$&--0.040(1)&--0.041(1)&S&--0.33(3)&i)&0.7974(8) \\
&$E_2$&0.050(1)&0.048(2)&0.053(4)&0.050(5)&ii)&0.7421(6) \\
&$E_3$&0.443(6)&0.41(3)&&&& \\ \hline 
(3,3)&$E_1$&--0.026(1)&--0.028(1)&S&S&i)&0.7974(8) \\
&$E_2$&"&"&--0.006(4)&--0.006(4)&&\\
&$E_3$&0.367(7)&0.366(6)&&&& \\ \hline 
(3,4)&$E_1$&--0.009(1)&--0.010(2)&0.025(9)&--0.008(1) 
&i)&0.7974(8)\\
&$E_2$&0.106(1)&0.105(5)&0.105(5)&0.12(1)&ii)&0.8589(11) \\
      &$E_3$&0.400(6)&0.39(3)&&&& \\ \hline
(4,3) &$E_1$&--0.036(2)&--0.039(2)&S&--0.016(2)&i)&0.9102(15) \\
      &$E_2$&0.033(2)&0.03(1)&0.03(2)&0.04(1)&ii)&0.8589(11) \\
      &$E_3$&0.326(6)&0.34(1)&&&& \\ \hline
(4,4) &$E_1$&--0.028(3)&--0.033(3)&S&S&i)&0.9102(15) \\
      &$E_2$&"&"&0.021(2)&0.021(2)&&\\
      &$E_3$&0.270(9)&0.26(1)&&&& \\ \hline
(4,5) &$E_1$&--0.009(3)&--0.012(2)&0.089(4)&--0.008(5)
&i)&0.9102(15) \\
      &$E_2$&0.089(3)&0.089(4)&0.089(5)&0.13(1)&ii)&0.967(2) \\
      &$E_3$&0.298(7)&0.31(1)&&&& \\ \hline
(5,4) &$E_1$&--0.025(5)&--0.04(1)&S&0.03(2)&i)&1.017(2) \\
      &$E_2$&0.025(5)&0.023(5)&0.05(1)&0.02(3)&ii)&0.967(2) \\
      &$E_3$&0.24(1)&0.24(1)&&&& \\ \hline
(5,5) &$E_1$&--0.017(8)&--0.10(4)&S&S&i)&1.017(2) \\
      &$E_2$&"&"&0.044(8)&0.44(8)&&\\
      &$E_3$&0.19(1)&0.19(1)&&&& \\ \hline
(5,6) &$E_1$&--0.06(3)&--0.013(4)&0.00(6)&--0.011(4) &i)&1.017(2) \\
      &$E_2$&0.01(5)&0.0(1)&0.0(1)&0.03(7)&ii)&1.072(4)\\
      &$E_3$&0.15(9)&0.14(10)&&&& \\
\end{tabular}
\end{center}
\end{table}

\begin{table}[p]
\caption{The notation is the same as Table \protect\ref{tetresults1}
but with $r=1$ and $d=2,3,4,5$ \protect\\ These results are from
3008 measurements contained in 47 blocks of 64 measurements each.
\protect\\ The $E_i(d\times d)$ are the corresponding energies for
the nearby square of side $d$. \protect\\ \label{tetresults2}}

%\renewcommand{\baselinestretch}{1.0}
%\Huge
%\large
\begin{center}
\begin{tabular}{cc|c|c|c||r@{\hspace {0.1cm}}l}
$(d/a,r/a)$&&$A$+$B$+$C$ (I)&$A$+$B$+$C$ (II) &$E_i(d\times d)$&
\multicolumn{2}{c}{$V_1$}\\ \hline
(2,1) &$E_1$&--0.0447(6)&--0.043(2)&--0.0588(3)&i)&0.6689(4) \\
      &$E_2$&0.085(1)&0.085(1)&0.1414(8)&ii)&0.6021(3)\\
      &$E_3$&0.69(3)&0.66(3)&&& \\ \hline
(3,1) &$E_1$&--0.052(2)&--0.049(2)&--0.0531(5) &i)&0.7974(8)  \\
      &$E_2$&0.089(3)&0.09(1)&0.1157(12)&ii)&0.6992(5)\\
      &$E_3$&0.53(2)&0.55(5)&&& \\ \hline
(4,1) &$E_1$&--0.050(4)&--0.047(1)&--0.0524(10)&i)&0.9102(15) \\
      &$E_2$&0.088(6)&0.088(3)&0.097(2)&ii)&0.7841(10)\\
      &$E_3$&0.54(5)&0.54(3)&&& \\ \hline
(5,1) &$E_1$&--0.052(5)&--0.044(5)&--0.047(3)&i)&1.017(2) \\
      &$E_2$&0.073(8)&0.073(1)&0.075(3)&ii)& 0.8652(8)\\
      &$E_3$&0.51(9)&0.55(4)&&& \\
\end{tabular}
\end{center}
\end{table}

\begin{table}
\caption{The energies for a selection of squares and rectangles in
the ranges $d=1,\ldots,5$ and $r=1,\ldots,5$. \protect\\ The $E_i$
are from a 2*2 basis using 800(1600) for the rectangles(squares). In
contrast to Ref.\ \protect\cite{GMS}, these results utilize Eqs.\
(\protect\ref{chi2})--(\protect\ref{Wex}) and not Eq.\
(\protect\ref{Eig}). \protect\\ The other columns use only 64
measurements. \protect\\ $A+B+C$ denotes the 3*3 basis and $A+B,\ \
B+C$ the corresponding 2*2 bases. \protect\\ The energies are those
at $T=3$ and not from any plateau in $T$. \protect\\
The notation is that:     \protect\\
$A$ has links along the sides of length $r$.  \protect\\
$B$ has links along the sides of length $d$.  \protect\\
$C$ is the state defined by the diagonals.    \protect\\
$S(L)$ is the state(s) with the lowest unperturbed energy. 
\protect\\ \label{t:square}}

\newpage 
\renewcommand{\baselinestretch}{1.0}
\Huge
\large
\begin{center}
\begin{tabular}{c|c|c|c|c|c|c}
$(d/a,r/a)$&&$E_i(P)$&$A+B+C$&$A+B$&$A+C$&$S(L)$ \\ \hline 
(1,1)&$E_1$&--0.0694(5)&--0.0694(3)&S&S &\\
&$E_2$&0.1828(7)&0.1841(11)&0.1842(11)&0.1846(11) &$A,B$\\
&$E_3$&-&0.974(25)&--&-- &\\ \hline

(1,2)&$E_1$&--0.0029(2)&--0.00150(18)&--0.00148(17)&--0.00075(16)
&\\ &$E_2$&0.504(2)&0.5074(19)&0.5077(20)&0.5089(20) &$B$\\
&$E_3$&-&1.037(19)&--&-- &\\ \hline
(2,1)&$E_1$&--0.0029(2)&--0.00165(38)&--0.00163(38)&--0.00164(38)
&\\ &$E_2$&0.504(2)&0.5080(28)&0.5083(29)&0.511(3) &$A$\\
&$E_3$&-&1.027(26)&--&-- &\\ \hline
(2,2)&$E_1$&--0.0588(3)&--0.0574(13)&S&--0.0566(14) &\\
&$E_2$&0.1414(8)&0.1429(14)&0.1433(13)&0.1496(7) &$A,B$\\
&$E_3$&-&0.766(18)&--&-- &\\ \hline
(2,3)&$E_1$&--0.0052(4)&--0.0038(11)&--0.0038(11)&+0.0019(14) &\\
&$E_2$&0.325(1)&0.324(4)&0.324(4)&0.329(4) &$B$\\
&$E_3$&-&0.807(12)&--&-- &\\ \hline
(3,2)&$E_1$&--0.0052(4)&--0.0044(3)&--0.0043(3)&--0.0044(2) &\\
&$E_2$&0.325(1)&0.326(3)&0.326(3)&0.342(3) &$A$ \\
&$E_3$&-&0.776(17)&--&-- &\\ \hline
(3,3)&$E_1$&--0.0531(5)&--0.0538(16)&S&--0.0493(19) &\\
&$E_2$&0.1157(12)&0.1113(27)&0.1112(27)&0.130(4) &$A,B$\\
&$E_3$&-&0.661(19)&--&-- &\\ \hline
(3,4)&$E_1$&--0.0041(6)&--0.0039(17)&--0.0039(17)&+0.018(3) &\\
&$E_2$&0.253(1)&0.255(15)&0.255(15)&0.260(14) &$B$\\
&$E_3$&-&0.705(29)&--&-- &\\ \hline
(4,3)&$E_1$&--0.0041(6)&--0.0070(22)&--0.0071(22)&--0.0072(22) &\\
&$E_2$&0.253(1)&0.261(7)&0.262(7)&0.290(4) &$A$ \\
&$E_3$&-&0.73(5)&--&-- &\\ \hline
(4,4)&$E_1$&--0.0524(10)&--0.0466(56)&S&--0.032(5) &\\
&$E_2$&0.097(2)&0.087(12)&0.088(12)&0.124(16) &$A,B$\\
&$E_3$&-&0.59(7)&--&-- &\\ \hline
(5,5)&$E_1$&--0.047(3)&--0.026(10)&S&--0.001(12) &\\
&$E_2$&0.075(3)&0.088(10)&0.092(12)&0.136(13) &$A,B$\\
&$E_3$&-&0.87(5)&--& &\\
\end{tabular}
\end{center}
\end{table}

\end{document}